\documentclass{moriond}

\def\be{\begin{equation}}
\def\ee{\end{equation}}
\def\bea{\begin{eqnarray}}
\def\eea{\end{eqnarray}}

\usepackage{cite}

\newcommand{\Photo}{}

\begin{document}
\vspace*{4cm}
\title{TOP PAIR PRODUCTION  WITH A JET WITH 
NLO QCD \\ OFF-SHELL EFFECTS \footnote{\it Presented at the $51^{st}$
    Rencontres de Moriond session devoted to QCD and High Energy
    Interactions, La Thuile, 19-26 March 2016.} \footnote{Preprint
    number: 
TTK-16-15.}} 

\author{MA\L{}GORZATA WOREK } 

\address{Institute for Theoretical Particle Physics and Cosmology\\
RWTH Aachen University, D-52056 Aachen, Germany}

\maketitle\abstracts{A brief summary of the recent next-to-leading
order QCD calculation for $e^+\nu_e \mu^- \bar{\nu}_\mu b\bar{b} j$ at
the LHC is given. Our computation includes all non-resonant
contributions, off-shell effects and interferences for top-quarks and
$W$ gauge bosons. Some results for integrated and differential cross
sections are shown for the LHC Run1 energy of 8 TeV. A significant
reduction of the scale dependence is observed, which indicates that
the perturbative expansion is well under control. The results are
obtained in the framework of the \textsc{Helac-Nlo} system.}

\begin{figure}
\begin{center}
\includegraphics[width=0.45\textwidth]{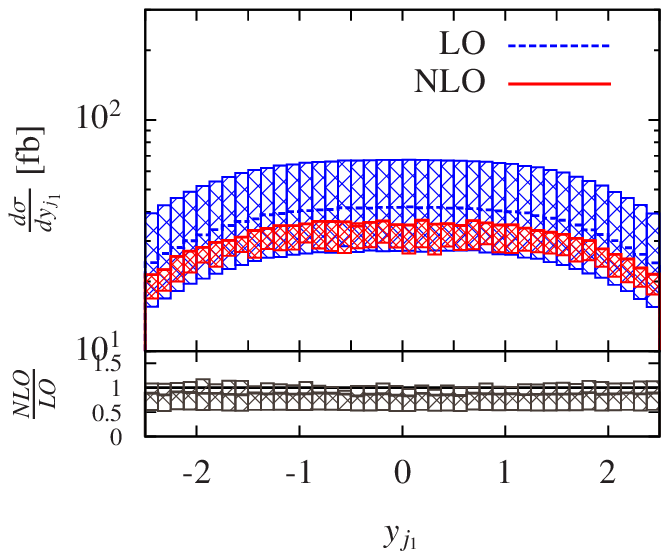}
\includegraphics[width=0.45\textwidth]{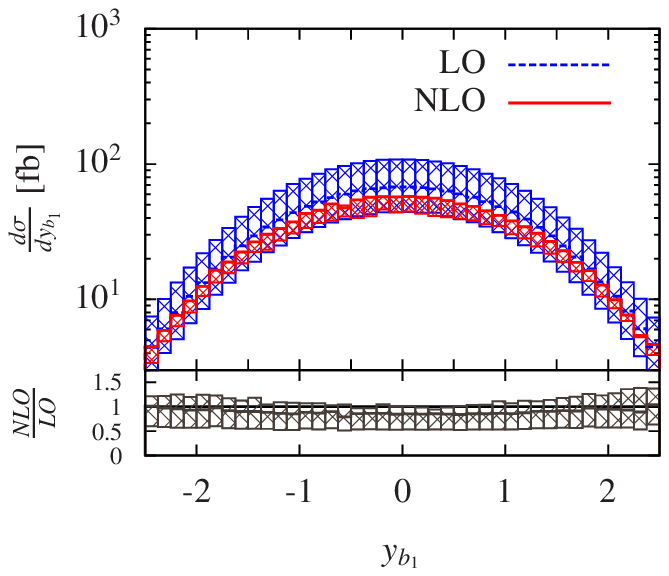}
\end{center}
\caption{\it   Rapidity of the hardest light- and   b-jet for
$pp\to e^+\nu_e \mu^- \bar{\nu}_\mu b\bar{b} j +X$ at the LHC with
$\sqrt{s} = 8 ~{\rm TeV}$.  The uncertainty bands depict the scale
variation. Lower panels display differential K factors and their 
uncertainty bands.}
\label{fig:distributions1}
\end{figure}

Since its discovery at the Tevatron till the collider's shut-down in
2011, the properties of the top quark and
its interactions have been studied in detail at the center-of-mass
energy of $\sqrt{s} = 1.96$ TeV.  These studies are now
continued at the LHC, which is in operation since the end of
2009. Starting from Run1 energies, i.e. $\sqrt{s}= 7, 8$ TeV and
continuing with Run2 energy of $\sqrt{s}= 13$ TeV many aspects of the
top quark physics have been examined very precisely.  Theoretical
predictions have also been significantly improved recently. By now
full ${\rm NNLO + NNLL}$ calculations for the total inclusive
$t\bar{t}$ cross section exist \cite{Czakon:2013goa} along with the
${\rm NNLO}$ level predictions for various differential distributions
\cite{Czakon:2015owf,Czakon:2016ckf}. The synergy between the very precise
theoretical predictions and the LHC data allowed to improve our
knowledge of the strong coupling constant and the top-quark mass,
which are both crucial parameters of the SM. Moreover, $\sigma^{\rm
NNLO+NNLL}_{pp\to t \bar{t}}$ theoretical predictions helped to
constrain the gluon parton distribution functions at large $x$, that
are crucial when calculating any cross sections in $pp$
collisions. Besides its tremendous role in improving our understanding
of QCD and the electroweak theory, the top quark plays an important
role in many scenarios for new physics beyond the SM, which
constitutes one of the main motivation for the top quark physics
program at the LHC.  Precise predictions for the $t\bar{t}$ cross
section helped to constrain BSM physics either by putting new
stringent limits on various new physics scenarios or by proposing new
ideas to improve search methods.  The large collision energy and
luminosity of the LHC, result in top quarks being produced in very
large quantities. Consequently, they are produced with large energies
and high transverse momenta, which increases the probability for
additional (hard) jet radiation and result in more exclusive final
states like for example $pp \to t\bar{t}j$ production. In order to
improve our knowledge of the inclusive $t\bar{t}$ cross section such
an exclusive final state must be well under control. The first
question that can arise is about the size of the $t\bar{t}j$
contribution to the inclusive $t\bar{t}$ sample. For a $p_T(j)$ cut
of $40$ GeV almost $40\%$ of $t\bar{t}$ events are actually
accompanied by an additional hard jet. A good understanding of the
$t\bar{t}j$ process is, thus, a prerequisite for a more precise
understanding of the topology of top-quark events. However, the $pp
\to t\bar{t}j$ process is also interesting by itself. It constitutes
the dominant background process to the Higgs boson production in the
vector boson fusion with Higgs boson decays into $W^+W^- \to 2\ell \,2
\nu $. Typical vector boson cuts for two tagging (hardest in $p_T$)
jets, denoted by $j_1, j_2$, consist of $\Delta y_{j_{1}j_{2}} =
|y_{j_1} -y_{j_2}| > 4$ and $y_{j_1} \cdot y_{j_2}<0$.  When comparing
rapidity distributions of the hardest light- and b-jet for two
production processes $pp \to t\bar{t} \to W^+W^- b\bar{b}$ and $pp \to
t\bar{t}j \to W^+W^-b\bar{b} j$ it is clearly visible that $b$-jets
are produced centrally while light-jets are distributed more evenly
(see Figure \ref{fig:distributions1}).  Asking for two tagging (b-)
jets to fulfil such requirements in case of $t\bar{t}$ will
dramatically decrease the contribution from the process. On the other
hand, for the $t\bar{t}j$ process in presence of the additional
light-jet it is sufficient that only one $b$-jet is considered to be
the tagging jet. As a consequence, not the inclusive $t\bar{t}$
production, but $t\bar{t}j$ is the dominant background process for $pp
\to H jj \to W^+W^- jj \to 2\ell \, 2 \nu jj$. In addition,
$t\bar{t}j$ production plays a very important role in searches for
physics beyond the SM. With $\ell$+jet and $\ell\ell$ final states
$t\bar{t}j$ is the main background to processes such as supersymmetric
particle production, where depending on the specific model, typical
signals include jets, charged leptons, and missing $p_T$ due to the
escaping lightest supersymmetric particle.
%
%
\begin{figure}
\begin{center}
\includegraphics[width=1.0\textwidth]{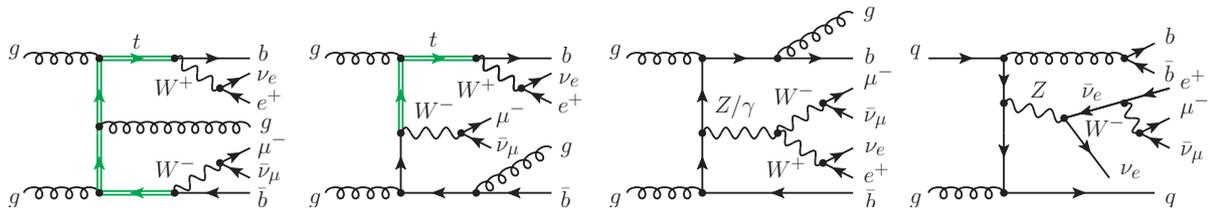}
\end{center}
\caption{\it Representative Feynman diagrams, involving two (first
  diagram), one (second diagram) and no top-quark resonances (third
  diagram),  contributing to the  $pp \to e^+\nu_e \mu^-
  \bar{\nu}_\mu b\bar{b}j$ process at ${\cal O}(\alpha_s^3 \alpha^4)
  $. The last diagram with a single $W$ boson resonance contributes to
  the off-shell  effects of  $W$ gauge bosons.  }
\label{fig:fd}
\end{figure}
%

Owing to the importance of $pp\to t\bar{t}j$ production we calculate
NLO QCD corrections for this process including all non-resonant
diagrams, interferences, and off-shell effects of top-quarks and $W$
gauge bosons \cite{Bevilacqua:2015qha}.  In practice $\alpha_s$
corrections are evaluated to the following LO process $pp\to e^+\nu_e
\mu\bar{\nu}_\mu b\bar{b}j$ at ${\cal O}(\alpha^3_s
\alpha^4)$. Representative LO Feynman diagrams are shown in Figure
\ref{fig:fd}. For the inclusive cross section contributions from top
quark off-shell effects are formally suppressed by the top-quark width
($\Gamma_t/m_t \approx 1\%$)
\cite{Denner:2010jp,Bevilacqua:2010qb,Denner:2012yc,Denner:2015yca}. They
can, however, be strongly enhanced for more exclusive observables
\cite{AlcarazMaestre:2012vp}. Here, NLO QCD corrections have been calculated
with the \textsc{Helac-Nlo} Monte Carlo program
\cite{Bevilacqua:2011xh}. The virtual corrections have been obtained with
\textsc{Helac-1Loop} \cite{vanHameren:2009dr} and \textsc{CutTools}
\cite{Ossola:2007ax}, which are based on the
Ossola-Papadopoulos-Pittau reduction technique
\cite{Ossola:2006us}. The \textsc{OneLOop} program
\cite{vanHameren:2010cp} has been  used for the evaluation of the scalar
integrals. The process under consideration requires a special
treatment of unstable top-quarks, which is achieved within the complex
mass scheme.  The singularities from soft or collinear parton
emissions are isolated via subtraction methods for NLO QCD
calculations that are implemented in \textsc{Helac-Dipoles} 
\cite{Czakon:2009ss}.  Specifically, two independent
subtraction schemes have been  employed: the commonly used Catani-Seymour
dipole subtraction \cite{Catani:1996vz,Catani:2002hc}, and a fairly
new Nagy-Soper subtraction scheme \cite{Bevilacqua:2013iha}. The phase
space integration was performed with the multichannel Monte Carlo
generator \textsc{Kaleu} \cite{vanHameren:2010gg}.

\begin{figure}[t!]
\begin{center}
\includegraphics[width=0.47\textwidth]{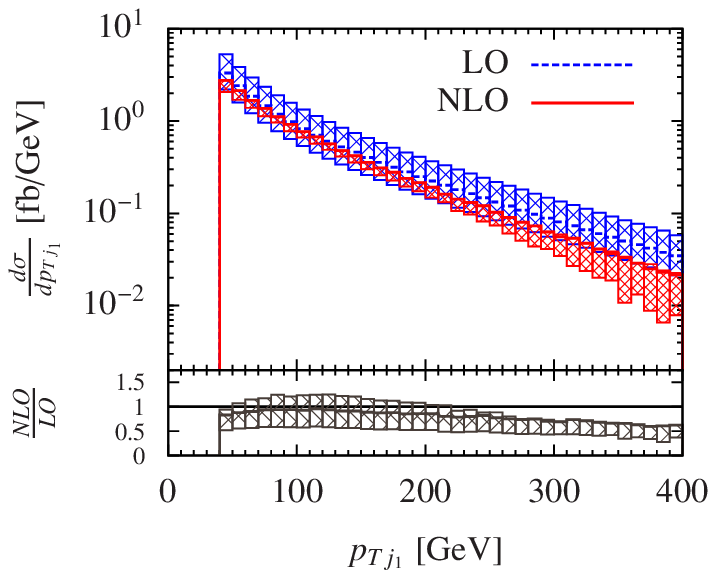}
\includegraphics[width=0.45\textwidth]{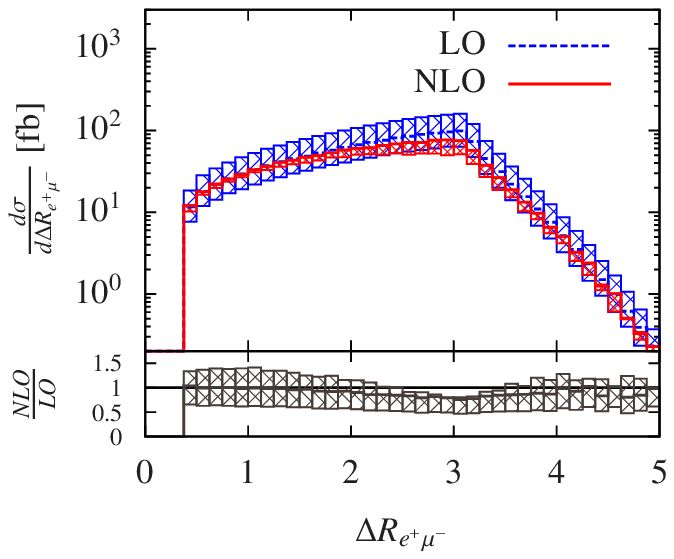}
\end{center}
\caption{\it   Transverse  momentum of the hardest light jet and  $\Delta
  R_{e^+\mu^-}$ for $pp\to e^+\nu_e \mu^-
  \bar{\nu}_\mu b\bar{b} j +X$  at the LHC with $\sqrt{s} = 8 ~{\rm
    TeV}$.  The uncertainty bands depict the scale variation. 
  Lower panels display differential K factors and their uncertainty bands.}
\label{fig:distributions2}
\end{figure}

In the following we present selected results for $pp \to e^+\nu_e \mu^-
\bar{\nu}_\mu b \bar{b} j+ X$ at the LHC with  $\sqrt{s} = 8$ TeV. The
SM parameters and cuts are specified  below 
%
\begin{table*}[h!]
\begin{center}
\begin{tabular}{| cc|cc| }
\hline
 $G_{\rm F}=1.16637 \cdot 10^{-5} ~{\rm GeV}^{-2}$ & $m_{\rm
  t}=173.3    ~{\rm GeV}$   & $p_{T \ell}>30 ~{\rm GeV}$ & 
$p_{Tj}>40 ~{\rm GeV}$ \\
$m_{\rm W}=80.399 ~{\rm GeV}$ & $\Gamma_{\rm W} = 2.09974 ~{\rm
                                  GeV}$ & $p^{\rm miss}_{T} >40 ~{\rm GeV}$ & $\Delta R_{jj}>0.5$\\
$m_{\rm Z}=91.1876  ~{\rm GeV}$
&$\Gamma_{\rm Z} = 2.50966 ~{\rm
  GeV}$ &$\Delta R_{\ell\ell}>0.4 $  &
 $\Delta R_{\ell j}>0.4$ \\
$\Gamma_{\rm t}^{\rm LO} = 1.48132 ~{\rm GeV}$ &
$\Gamma_{\rm
  t}^{\rm NLO} = 1.3542 ~{\rm  GeV}$ 
& $|y_\ell|<2.5$ & $|y_j|<2.5$\\
\hline
\end{tabular}
 \end{center}
\end{table*}
%
where $\ell$ stands for $\mu^-,e^+$ and $j$ for the light- or $
b$-jet. Jets are defined by the anti-$k_{\rm T}$ jet algorithm with
the separation parameter $R=0.5$ and MSTW2008 parton distribution
functions are  chosen.  Results for the total cross sections
are as follows
\begin{equation}
\sigma^{\rm LO} 
= 183.1^{\, +112.2 \, (61\%)}_{\,\,\,\,  -64.2 \, (35\%)} ~{\rm fb}\,,
~~~~~~~~~~~~
\sigma^{\rm NLO}
= 159.7^{\, -33.1 \,(21\%)}_{\,\,\,\, -7.9 \, (\,\,\, 5\%)} ~ {\rm
  fb}\,, ~~~~~~~~~~~~
{\cal K} = \sigma^{\rm NLO}/\sigma^{\rm LO} =0.87 \,.
\end{equation}
The full $pp$ cross section receives negative and moderate NLO
corrections of $13\%$ at the central scale, i.e. for $\mu=\mu_{\rm
R}=\mu_{\rm F}=m_{\rm t}$.  Theoretical uncertainties, associated with
neglected higher order terms in the perturbative expansion, have been
estimated to be $61\%$ at LO and $21\%$ at NLO. Thus, a reduction of the
theoretical error by a factor of 3 was observed. We have also assessed
the size of the non-factorizable corrections. At LO (NLO) finite
top-quark width effects changed the cross section by less than $1\%
\,(2\%)$. Representative differential distributions are presented in
Figure \ref{fig:distributions2}, where we exhibit the transverse
momentum of the hardest (in $p_T$) light jet and the separation
between charged leptons in the rapidity azimuthal angle plane.  The
dashed (blue) curve corresponds to the LO, whereas the solid (red) one
to the NLO result. The upper panels show the distributions themselves
and the scale dependence bands. The lower panels display the
differential ${\cal K}$ factor. Higher order corrections to $p_{Tj_1}$
do not simply rescale the shape of the LO distribution. 
Corrections up to $50\%$ are introduced away from the threshold for
the $t\bar{t}j$ production.  Thus, the $p_{Tj_1}$ differential
cross section can only be properly described when the higher order
corrections are taken into account.  A judicious choice of the dynamic
scale, could, however, change negative NLO corrections in the high
$p_T$ tails and a constant ${\cal K}$ factor could be achieved 
in the whole $p_T$ region. On the contrary, for the $\Delta R_{e^+\mu^-}$
distribution, negative, moderate and quite stable corrections have
been observed,  because $d\sigma/d\Delta R_{e^+\mu^-}$ receives
contributions from all scales, most notably from those that are
sensitive to the threshold for the $t\bar{t}j$ production. Indeed, for
our scale choice, effects of the phase space regions close to this
threshold dominate and a dynamic scale will not alter the behaviour in
that case.

To summarise, we have calculated NLO QCD corrections to $pp \to
e^+\nu_e\mu^-\bar{\nu}_\mu b\bar{b}j+X$ with complete off-shell and
interference effects both for top-quarks and $W$ gauge bosons. We have
shown that NLO QCD corrections to the total cross section are moderate
but their impact on some differential distributions is much larger. We
have also estimated the size of the top quark off-shell effects at NLO
for the total cross section, and confirmed that they are of the order
of ${\cal O}(\Gamma_{\rm t}/m_{\rm t})$. Let us stress here, that
$t\bar{t}j$ process can add to alternative methods of determination of
the top-quark mass. One method recently proposed involves the
invariant mass of the $t\bar{t}j$ system \cite{Alioli:2013mxa,
Aad:2015waa}.  However, to extract the top-quark mass as precisely as
possible the most complex calculation for $t\bar{t}j$ need to be
considered that consists of a full simulation of the final state
without any approximations.  Thus, in the next step our results can be
used to extract  the top-quark pole mass with a very high precision.

The author would like to thank the organizers of Recontres de Moriond
for the kind invitation and the very pleasant atmosphere during the
conference. The work was supported by the DFG under Grant: {\it "Signals and
Backgrounds Beyond Leading Order. Phenomenological studies for the
LHC".}

%
%

\end{document}